\documentclass[prd,aps,preprintnumbers,floats,floatfix,superscriptaddress,preprintnumbers,
showpacs,eqsecnum,nofootinbib,twocolumn
]{revtex4-1}
\usepackage[utf8]{inputenc}
\usepackage{latexsym,array,theorem,mathrsfs,bm,float}
\usepackage{psfrag}
\usepackage{amsfonts,amsmath,amssymb,latexsym,array,afterpage,
theorem,mathrsfs,bm,float,epsfig,color,graphicx,tabularx,here,multirow}

\newcommand{\nn}{\nonumber \\}
\newcommand{\bea}{\begin{eqnarray}}
\newcommand{\ena}{\end{eqnarray}}
\newcommand{\beann}{\begin{eqnarray*}}
\newcommand{\enann}{\end{eqnarray*}}

\begin{document}
\baselineskip=12pt

\preprint{YITP-17-81} 
\preprint{IPMU17-0106} 
\preprint{WU-AP/1703/17}

\title{Massive graviton dark matter with environment dependent mass: \\
A natural explanation of the dark matter-baryon ratio
}
\author{Katsuki \sc{Aoki}}
\email{katsuki-a12@gravity.phys.waseda.ac.jp}
\affiliation{
Department of Physics, Waseda University,
Shinjuku, Tokyo 169-8555, Japan
}

\author{Shinji \sc{Mukohyama}}
\email{shinji.mukohyama@yukawa.kyoto-u.ac.jp}
\affiliation{Center for Gravitational Physics, Yukawa Institute for Theoretical Physics, Kyoto University, 606-8502, Kyoto, Japan}
\affiliation{Kavli Institute for the Physics and Mathematics of the
Universe (WPI), UTIAS, The University of Tokyo, Kashiwa, Chiba
277-8583, Japan}

\date{\today}

\begin{abstract}
We propose a scenario that can naturally explain the observed dark matter-baryon ratio in the context of bimetric theory with a chameleon field. We introduce two additional gravitational degrees of freedom, the massive graviton and the chameleon field, corresponding to dark matter and dark energy, respectively. The chameleon field is assumed to be non-minimally coupled to dark matter, i.e., the massive graviton, through the graviton mass terms. We find that the dark matter-baryon ratio is dynamically adjusted to the observed value due to the energy transfer by the chameleon field. As a result, the model can explain the observed dark matter-baryon ratio independently from the initial abundance of them.
\end{abstract}



\maketitle



\section{Introduction}
Cosmological observations have confirmed the big bang cosmology and determined the cosmological parameters precisely~\cite{Ade:2015xua}. The matter contents of the Universe may be phenomenologically given by the standard model particles, the cosmological constant $\Lambda$, and cold dark matter (CDM). However, the theoretical explanation of the origin of the extra ingredients, dark matter and dark energy, is still lacked. The theoretically expected value of the cosmological constant is too large to explain the present accelerating expansion. An alternative idea is that the acceleration is obtained by a potential of a scalar field instead of $\Lambda$, and this idea is often called the quintessence model~\cite{Caldwell:1997ii}. This scalar field could be originated from the gravity sector~\cite{Fujii:2003pa}. A large class of scalar-tensor theories and $f(R)$ theories can be recast in the form of a theory of a canonical scalar field with a potential after the conformal transformation $\tilde{g}_{\mu\nu}=A^2(\phi) g_{\mu\nu}$ and the field redefinition $\Phi=\Phi(\phi)$ where $\Phi$ is the canonically normalized field. The metric $\tilde{g}_{\mu\nu}$ is called the Jordan frame metric which the standard model particles are minimally coupled to whereas $g_{\mu\nu}$ is the Einstein frame metric in which the gravitational action is given by the Einstein-Hilbert action. In this case, the scalar field has the non-minimal coupling to the matter fields via the coupling function $A$.

Dark matter is also one of the biggest mystery of the modern cosmology. Although many dark matter candidates have been proposed in the context of the particle physics, any dark matter particles have not been discovered yet~\cite{Agashe:2014kda,Ackermann:2015zua,Ahnen:2016qkx,Ackermann:2015lka,Khachatryan:2014rra,Conrad:2017pms}. The existence of dark matter is confirmed via only gravitational interactions. Hence, exploring dark matter candidate in the context of gravity is also a considerable approach. Not only dark energy but also dark matter could be explained by modifications of gravity. For instance, a natural extension of general relativity is a theory with a massive graviton (see \cite{deRham:2014zqa} for a review). If a graviton obtains a mass, the massive graviton can be a dark matter candidate~\cite{Dubovsky:2004ud,Pshirkov:2008nr,Aoki:2016zgp,Babichev:2016hir,Babichev:2016bxi,Aoki:2017cnz}.

A viable dark matter scenario has to explain the present abundance of dark matter which usually leads to a constraint on a production scenario. However, a question arises: why are the energy densities of dark matter and baryon almost the same? If baryon and dark matter are produced by a common mechanism, almost the same abundance could be naturally obtained. On the other hand, if productions of the two are not related but independent, the coincidence might indicate that two energy densities are tuned to be the same order of the magnitude by a mechanism after the productions.

In the present paper, we shall combine two ideas of the modifications of gravity by using the proposal of \cite{DeFelice:2017oym}: the non-minimal coupling of $\phi$ and the existence of the massive graviton. We call this theory the chameleon bigravity theory which contains three types of gravitational degrees of freedom: the massless graviton, the massive graviton, and the chameleon field $\phi$. We identify the massive graviton with dark matter. Since dark matter is originated from the gravity sector, the coupling between $\phi$ and the dark matter may be given by a different way from the matter sector. We promote parameters in the graviton mass terms to functions of $\phi$~\cite{D'Amico:2012zv,Huang:2012pe}, giving rise to a new type of coupling between $\phi$ and dark matter. In this case, as discussed in \cite{DeFelice:2017oym}, the field value of $\phi$ depends on the environment due to the non-minimal coupling as with the chameleon field \cite{Khoury:2003aq,Khoury:2003rn}, which makes the graviton mass to depend on the environment.

We find that the ratio between energy densities of dark matter and baryon is dynamically adjusted to the observed value by the motion of $\phi$ and then the ratio at the present is independent of the initial value. Hence, our model can explain the coincidence of the abundance of dark matter and baryon. Furthermore, if the potential of $\phi$ is designed to be dark energy, the chameleon field $\phi$ can give rise to the present acceleration of the universe. Both dark energy and dark matter are explained by the modifications of gravity in our model.

The paper is organized as follows. We introduce the chameleon bigravity theory in Sec.~\ref{sec_bigravity}. In Sec.~\ref{sec_Fri}, we show the Friedmann equation regarding the massive graviton is dark matter. We also point out the reason why the dark matter-baryon ratio can be naturally explained in the chameleon bigravity theory if we consider the massive graviton as dark matter. Some analytic solutions are given in Sec.~\ref{sec_analytic} and numerical solutions are shown in Sec.~\ref{sec_numerical}. These solutions reveal that the observed dark matter-baryon ratio is indeed dynamically obtained independently from the initial ratio. We summarize our results and give some remarks in Sec. \ref{summary}. In Appendix, we detail the derivation of the Friedmann equation.

\section{Chameleon bigravity theory}
\label{sec_bigravity}

We consider the chameleon bigravity theory in which the mass of the massive graviton depends on the environment~\cite{DeFelice:2017oym}. The action is given by
\begin{align}
S&=\int d^4x \sqrt{-g} \Biggl[  \frac{M_g^2}{2}R[g]-\frac{1}{2}
 K^2(\phi) 
g^{\mu\nu}\partial_{\mu}\phi\partial_{\nu}\phi
\nn
&\qquad \qquad \qquad \quad +M_g^2m^2 \sum_{i=0}^4 \beta_i (\phi) U_i[s] \Biggl]
\nn
&+\frac{M_f^2}{2}\int d^4 x\sqrt{-f}R[f]+S_{\rm m}[\tilde{g},\psi]\,, \label{action}
\end{align}
where $\phi$ is the chameleon field and $S_{\rm m}$ is the matter action. The functions $K(\phi)$ and $\beta_i(\phi)$ are arbitrary functions of $\phi$. The matter fields universally couple to the Jordan frame metric $\tilde{g}_{\mu\nu}=A^2(\phi)g_{\mu\nu}$ with a coupling function $A(\phi)$. The potentials $U_i[s] \, (i=0,\cdots, 4)$ are the elementary symmetric polynomials of the eigenvalues of the matrix $s^{\mu}{}_{\nu}$ which is defined by the relation~\cite{deRham:2010ik,deRham:2010kj,Hassan:2011zd}
\begin{align}
s^{\mu}{}_{\alpha}s^{\alpha}{}_{\nu}=g^{\mu\alpha}f_{\alpha\nu}\,.
\end{align}
The potential of $\phi$ is not added explicitly since the couplings between $\phi$ and the potentials $U_i$ yield the potential of $\phi$ and thus an additional potential is redundant.

Note that the field $\phi$ is not a canonically normalized field. The canonical field $\Phi$ is given by the relation
\begin{align}
d \Phi = K(\phi) d\phi \,,
\end{align}
by which the function $K$ does not appear explicitly in the action when we write down the theory in terms of $\Phi$. Since $\beta_i$ and $A$ are arbitrary functions, we can set $K=1$ by the redefinitions of $\beta_i$ and $A$ without loss of generality. Nevertheless, we shall retain $K$ and discuss the general form of the action \eqref{action}.

In general, the functions $\beta_i(\phi)$ can be chosen independently. In the present paper, however, we consider the simplest model such that $\beta_i(\phi)=-c_i f(\phi)$ where $c_i$ are constant while $f(\phi)$ is a function of $\phi$. As we will see in next section, the graviton mass and the potential of $\phi$ around the cosmological background are given by
\begin{align}
m_T^2(\phi)&:=\frac{1+\kappa}{\kappa}m^2 f(\phi)(c_1+2c_2+c_3)
\,, \\
V_0(\phi)&:=m^2M_p^2 f(\phi)(c_0+3c_1+3c_2+c_3)
\,, \label{bare_potential}
\end{align}
with $\kappa=M_f^2/M_g^2$ and $M_p^2=M_g^2+M_f^2$. In this case, both the potential form of $\phi$ and the $\phi$-dependence of the graviton mass are determined by $f(\phi)$ only.\footnote{Since we have absorbed the potential of $\phi$ in the mass term of the graviton, $m_T^2M_p$ and $V_0$ seem to be a same order of magnitude. However, $m_T^2M_p^2$ and $V_0$ are not necessary to be the same order because they represent different physical quantities. Indeed, we will assume $V_0 \ll m_T^2 M_p^2$.} Note that $V_0$ is the bare potential of $\phi$. The effective potential of $\phi$ is given by not only $V_0$ but also the amplitude of the massive graviton as well as the energy density of matter due to the non-minimal couplings (see Eq.~\eqref{effective_potential}).

\section{Basic equations}
\label{sec_Fri}
In this section, we derive the basic equations to discuss the cosmological dynamics in the model \eqref{action} supposing that the massive graviton is dark matter. We assume the coherent dark matter scenario in which dark matter is obtained from the coherent oscillation of the zero momentum mode massive gravitons~\cite{Aoki:2017cnz}. Since the zero momentum mode of the graviton corresponds to the anisotropy of the spacetime, we study the Bianchi type I universe instead of the Friedmann-Lema{\^i}tre-Robertson-Walker (FLRW) universe. The ansatz of the spacetime metrics are
\begin{align}
ds_g^2&=-dt^2 +a^2[ e^{4\sigma_g} dx^2+e^{-2\sigma_g}(dy^2+dz^2)]\,, \label{Bianchi_g} \\
ds_f^2&=\xi^2\left[ -c^2 dt^2+a^2\{ e^{4\sigma_f} dx^2+e^{-2\sigma_f}(dy^2+dz^2)\} \right] \,, \label{Bianchi_f}
\end{align}
where $\{a,\xi,c,\sigma_g,\sigma_f\}$ are functions of the time $t$. We assume the matter field is a perfect fluid whose energy-momentum tensor is given by
\begin{align}
T^{\mu}{}_{\nu}=A^4(\phi)\times {\rm diag}[-\rho(t),P(t),P(t),P(t)]\,, \label{Tmunu}
\end{align}
where $\rho$ and $P$ are the energy density and the pressure in the Jordan frame, respectively. The conservation law of the matter field is
\begin{align}
\dot{\rho}+3\frac{(Aa)^{\cdot}}{Aa}(\rho+P)=0\,, \label{conservation}
\end{align}
where a dot is the derivative with respect to $t$.

As shown in \cite{Maeda:2013bha,Aoki:2017cnz}, the small anisotropies $\sigma_g$ and $\sigma_f$ can be a dark matter component of the universe in the bimetric model without the chameleon field $\phi$. We generalize their calculations to those in the present model \eqref{action}. All equations under the ansatz \eqref{Bianchi_g} and \eqref{Bianchi_f} are summarized in Appendix. Here, we only show the Friedmann equation and the equations of motion of the massive graviton and the chameleon field because other equations are not important for the following discussion.

We assume the graviton mass $m_T$ is larger than the Hubble expansion rate $H:=\dot{a}/a$. After expanding the equations in terms of anisotropies and a small parameter $\epsilon:=H/m_T$, the Friedmann equation is given by
\begin{align}
3M_p^2H^2= \rho A^4+\frac{1}{2}
 K^2 
\dot{\phi}^2+V_0+\frac{1}{2}\dot{\varphi}^2+\frac{1}{2}m_T^2\varphi^2 \,, \label{Fri_no_h}
\end{align}
where $\varphi$ is the massive graviton which is given by a combination of the anisotropies $\sigma_g$ and $\sigma_f$ (see Eqs.~\eqref{graviton1} and \eqref{graviton2}).
The equations of motion of the massive graviton $\varphi(t)$ and the chameleon field $\phi(t)$ are 
\begin{align}
\ddot{\varphi}+3H \dot{\varphi}+m_T^2(\phi) \varphi &=0 \,, \label{eq_varphi}\\
 K \left( \ddot{\phi}+3H\dot{\phi} \right)
 + \dot{K} \dot{\phi} +  \frac{\partial V_{\rm eff}}{ \partial \phi}&=0\,, \label{eq_cham}
\end{align}
where the effective potential of the chameleon field is given by
\begin{align}
V_{\rm eff}:= V_0(\phi)+\frac{1}{2}m_T^2(\phi)\varphi^2 + \frac{1}{4}A^4(\phi) (\rho-3p)
\,. \label{effective_potential}
\end{align}
Note that, although the bigravity theory contains the degree of freedom of the massless graviton (see Eq.~\eqref{Friedmann_eq}), we neglect the contribution to the Friedmann equation from the massless graviton because the energy density of the massless graviton decreases faster than those of other fields. The effect of the massless graviton is not important for our discussions.

We notice that the basic equations \eqref{Fri_no_h}, \eqref{eq_varphi} and \eqref{eq_cham} are exactly the same as the equations in the theory with two scalar fields given by the action
\begin{align}
S=\int d^4x \sqrt{-g} \Biggl[ &\frac{M_p^2}{2} R[g]
-\frac{1}{2}K^2(\phi) (\partial \phi)^2 -V_0(\phi) \nn
&-\frac{1}{2} (\partial \varphi)^2 -\frac{1}{2}m_T^2(\phi) \varphi^2 \Biggl]+S_m [\tilde{g},\psi]
\,. \label{another_action}
\end{align}
The cosmological dynamics in \eqref{action} with $H \gg m_T$ can be reduced into that in \eqref{another_action}. Our results obtained below can be straightforward generalized even in the case of \eqref{another_action} up to the discussion about the cosmological dynamics. The action \eqref{another_action} gives a toy model of the chameleon bigravity theory. However, the equivalence between \eqref{action} and \eqref{another_action} holds only for the background dynamics of the universe in $H\gg m_T$. The equivalence between the two actions does not hold for small-scale perturbations around the cosmological background~\cite{Aoki:2017cnz}.

We first consider a solution $\phi=\phi_{\rm min}=$ constant which is realized when
\begin{align}
\frac{\partial V_{\rm eff}}{\partial \phi}
=\alpha_f \left[ V_0 +\frac{1}{2}m_T^2 \varphi^2 \right] +\alpha_A (\rho-3P) A^4=0
\,,
\label{phi=const}
\end{align}
where
\begin{align}
\alpha_A :=\frac{1}{K}\frac{ d \ln A}{d \phi}=\frac{ d \ln A}{d \Phi}\,, \quad
\alpha_f:=\frac{1}{K}\frac{d \ln f}{d \phi}=\frac{ d \ln f}{d \Phi} \,.
\end{align}
The equation \eqref{phi=const} is not always compatible with $\phi=$ constant since each term in \eqref{phi=const} has different time dependence in general. Nonetheless, as we shall see below, they can be compatible with each other if $\epsilon \ll 1 $. In other words, a common constant value of $\phi_{\rm min}$ can be a solution all the way from the radiation dominant (RD) epoch to the matter dominant (MD) epoch of the universe. When the chameleon field is constant, the bare potential $V_0$ acts as a cosmological constant which has to be subdominant in the RD and the MD eras. The constant $\phi$ implies that the graviton mass does not vary and thus we obtain
\begin{align}
\langle \dot{\varphi}^2 \rangle_T= \langle m_T^2 \varphi^2 \rangle_T \propto a^{-3}
\,,\label{eqn:scaling-mTconst}
\end{align}
where $\langle\cdots \rangle_T$ represents the time average over an oscillation period. The massive gravitons behave like a dark matter component of the universe. When we focus on the time scales much longer than $m_T^{-1}$, $m_T^2 \varphi^2$ in Eq.~\eqref{phi=const} can be replaced with $\langle m_T^2 \varphi^2 \rangle_T$, which scales as \eqref{eqn:scaling-mTconst}. Since $\rho-3P$ also scales as $\propto a^{-3}$ in the RD and the MD, the decaying laws of $\rho-3P$ and $m_T^2 \varphi^2$ in \eqref{phi=const} are the same in this case. Hence, when the oscillation timescale of the massive graviton is much shorter than the timescale of the cosmic expansion, i.e., $\epsilon \ll 1 $, $\phi=$ constant can be a solution all the way from the RD to the MD. The value of $\phi_{\rm min}$ is determined by simply solving Eq.~\eqref{phi=const}.

Supposing that the massive graviton is the dominant component of dark matter, Eq.~\eqref{phi=const} in the RD and MD eras is replaced with
\begin{align}
\left( \alpha_f \rho_G+2\alpha_A \rho_{b} \right) A^4=0\,,
\end{align}
where $\rho_b$ is the baryon energy density and we have ignored $V_0$. The energy density of massive graviton in the Jordan frame is defined by
\begin{align}
\rho_G:=\frac{1}{2}A^{-4}\langle \dot{\varphi}^2+m_T^2\varphi^2 \rangle_T=A^{-4}m_T^2\langle \varphi^2 \rangle_T\,,
\end{align}
which depends on the chameleon $\phi$. Therefore, if $\alpha_A$ and $\alpha_f$ are assumed to be $\alpha_A/\alpha_f \simeq -5/2$, the ratio between dark matter and baryon is automatically tuned to be the observational value. The dark matter-baryon ratio could be naturally explained without any fine-tuning of the productions of dark matter and baryon.

Needless to say, the initial value of $\phi$ must not be at the bottom of the effective potential $(\phi=\phi_{\rm min})$. We shall study the dynamics of $\phi$ and discuss whether $\phi$ approaches $\phi_{\rm min}$ before the MD era of the universe. 
Although we do not assume $\phi$ is constant, we assume $\phi$ does not rapidly move so that the graviton mass varies adiabatically
\begin{align}
\frac{\dot{m}_T}{m_T^2} \ll 1 \,.
\label{adiabatic_condition}
\end{align}
Under the adiabatic condition  \eqref{adiabatic_condition} we can take the adiabatic expansion for the massive graviton:
\begin{align}
\varphi=u(t)\cos\left[  \int m_T[\phi(t)] dt \right]+\cdots\,,
\end{align}
with a slowly varying function $u(t)$. The adiabatic condition \eqref{adiabatic_condition} is indeed viable for $\epsilon \ll 1$ since we will see the time dependence of $m_T$ is given by a power law of $a$ (see Eq.~\eqref{time_dep_m} for example). The time average over an oscillation period yields $\langle \dot{\varphi}^2 \rangle_T=\langle m_T^2 \varphi^2 \rangle_T=m_T^2 u^2/2$.

After taking the time average over an oscillation period under the adiabatic condition, the equations are reduced into
\begin{align}
3M_p^2H^2= A^4 \rho_r +A^4 \rho_b +\frac{1}{2}  K^2  \dot{\phi}^2 +V_0+ \frac{1}{2}m_T^2 u^2 \,, \label{Fri}
\end{align}
and
\begin{align}
 K \left( \ddot{\phi}+3H \dot{\phi} \right)
 +\dot{K} \dot{\phi} +\alpha_f V_0 
& \nn
+\frac{1}{4}\alpha_f m_T^2 u^2+\alpha_A A^4\rho_b&=0
\,, \label{eq_phi} \\
4\dot{u}+6H u+ \alpha_f u K\dot{\phi}&=0
\,, \label{eq_u}
\end{align} 
where $\rho_r$ and $\rho_b$ are the energy densities of radiation and baryon which decrease as $\rho_r\propto (aA)^{-4}$ and $\rho_b \propto (aA)^{-3}$ because of the conservation equation. The dynamics of the scale factor $a$, the chameleon field $\phi$, and the amplitude of the massive graviton $u$ are determined by solving these three equations.

By using the density parameters, the Friedmann equation is rewritten as
\begin{align}
1=\Omega_r+\Omega_b+\Omega_{\phi}+\Omega_G \,,
\end{align}
with
\begin{align}
\Omega_r&:=\frac{A^4\rho_r}{3M_p^2H^2} 
\,, \\
\Omega_b&:=\frac{A^4\rho_b}{3M_p^2H^2} 
\,, \\
\Omega_{\phi}&:=\frac{\dot{\phi}^2+2V_0}{6M_p^2H^2}
\,, \\
\Omega_G&:=\frac{m_T^2 u^2}{6M_p^2H^2}\,.
\end{align}
We also introduce the total equation of state parameter in the Einstein frame
\begin{align}
w_E:=-1-\frac{2\dot{H}}{3H^2}\,.
\end{align}

The above quantities are defined in the Einstein frame. Since the matter fields minimally couple with the Jordan frame metric, the observable universe is expressed by the Jordan frame metric. Hence, we also define the Hubble expansion rate and the effective equation of state parameter in the Jordan frame as
\begin{align}
H_J&:=\frac{(Aa)^{\cdot}}{A^2a} \,, \\
w_{\rm tot}&:=-1-\frac{2\dot{H}_J}{3AH_J^2}\,.
\end{align}


\section{Analytic solutions}
\label{sec_analytic}
In this section we show some analytic solutions under the simplest case
\begin{align}
K=1\,, \quad A=e^{\beta\phi/M_p}\,, \quad
f=e^{-\lambda \phi/M_p}\,, \label{model_A}
\end{align}
with the dimensionless constants $\beta$ and $\lambda$. This model yields that the coupling strengths $\alpha_A$ and $\alpha_f$ are constant.
We consider four stages of the universe: the radiation dominant era, around the radiation-matter equality, the matter dominant era, and the accelerate expanding era. The analytic solutions are found in each stages of the universe as follows.

\subsection{Radiation dominant era}
We first consider the regime when the contributions to the Friedmann equation from baryon and dark matter are subdominant, that is, $\Omega_b, \Omega_G \ll 1$. The Hubble expansion rate is then determined by the energy densities of radiation and $\phi$. Since the effective potential of $\phi$ are determined by the energy densities of baryon and dark matter, in this situation the potential force can be ignored compared with the Hubble friction term ($V_0$ is assumed to be always ignored during both radiation and matter dominations).
Then, we obtain
\begin{align}
\dot{\phi}\propto a^{-3}
\,,
\end{align}
which indicates that the field $\phi$ loses its velocity due to the Hubble friction and then $\phi$ becomes a constant $\phi_i$. We can ignore $\Omega_{\phi}$ and then find the standard RD universe. At some fixed time deep in the radiation dominant era, we therefore set $\phi=\phi_i$ as the initial condition of $\phi$. We shall then denote the initial values of the energy densities of baryon and the massive graviton as $\rho_{b,i}$ and $\rho_{G,i}$, respectively.

Note that this constant initial value of $\phi$ is not necessary to coincide with the potential minimum $\phi=\phi_{\rm min}$, i.e., $\phi_i \neq \phi_{\rm min}$. The ratio $\rho_{G,i}/\rho_{b,i}$ is not tuned to be five at this stage.

\subsection{Following-up era}
We then discuss the era just before radiation-matter equality in which we cannot ignore the potential force for $\phi$. As discussed in the previous subsection, we find $\phi=\phi_i$ in the RD universe. When the potential force for $\phi$ becomes relevant, the chameleon field $\phi$ starts to evolve into the potential minimum $\phi=\phi_{\rm min}$. Due to the motion of $\phi$, the smaller one of $\rho_G$ and $\rho_b$ follows up the larger one. We obtain $\rho_G/\rho_b=-2\alpha_A/\alpha_f$ when the chameleon field reaches the minimum $\phi_{\rm min}$. We call this era of the universe the following-up era.

If the initial value $\phi_i$ is close to the potential minimum $\phi_{\rm min}$, the dark matter-baryon ratio is already tuned to be almost the value $-2\alpha_A/\alpha_f$, which we set to $\sim 5$, and thus we do not need to discuss this case. We therefore study the case with $\phi_i < \phi_{\rm min}$ and the case with $\phi_i > \phi_{\rm min}$ (which correspond to $\rho_{G,i} \gg \rho_{b,i}$ and $\rho_{b,i} \gg \rho_{G,i}$, respectively). We shall discuss them in order.

\subsubsection{$\rho_{G,i}\gg \rho_{b,i}$ before the equal time}
\label{DM>>b}
If dark matter (i.e., massive gravitons) is over-produced, the equations are reduced to
\begin{align}
\ddot{\phi}+3H \dot{\phi}-\frac{\lambda}{4M_p} m_T^2 u^2 =0\,,
\\
3M_p^2 H^2 =A^4 \rho_r +\frac{1}{2}\dot{\phi}^2+3m_T^2 u^2
\,,
\end{align}
and \eqref{eq_u}, where we have ignored the contributions from baryon. The system admits a scaling solution
\begin{align}
\phi&=\frac{M_p}{\lambda} \ln t +{\rm constant}\,,
\nn
u&\propto t^{-1/2}
\,,
\nn
a&\propto t^{1/2}
\,, \label{scaling_DM}
\end{align}
where the density parameters in the Einstein frame are given by
\begin{align}
\Omega_G=\frac{4}{3\lambda^2}
\,, \quad
\Omega_{\phi}=\frac{2}{3 \lambda^2}
\,, \quad
\Omega_r=1-\frac{2}{\lambda^2}
\,.
\end{align}
The effective equation of state parameter in the Jordan frame is given by
\begin{align}
w_{\rm tot}=\frac{\lambda-2\beta}{3(\lambda+2\beta)} \,,
\end{align}
and then $w_{\rm tot} =-2/9$ if $2\beta=5\lambda$. This solution exists only when $\lambda^2>2$ since the density parameter has to be $0<\Omega_r<1$.

For this scaling solution, the graviton mass decreases as
\begin{align}
m_T^2 \propto a^{-2}\,, \label{time_dep_m}
\end{align}
which guarantees the adiabatic condition \eqref{adiabatic_condition} when $\epsilon \ll 1$.
The energy density of massive gravitons in the Einstein frame decreases as
\begin{align}
A^4 \rho_G =\frac{1}{2}m_T^2 u^2 \propto a^{-4}
\,.
\end{align}
On the other hand, the energy density of baryon in the Einstein frame ``increases'' as
\begin{align}
A^4 \rho_b \propto A a^{-3} \propto a^{-3+2\frac{\beta}{\lambda}}
\,,
\end{align}
(For example, we obtain $A^4 \rho_b \propto a^2$ when $2\beta = 5 \lambda$). Therefore, even if baryon is negligible at initial, the baryon energy density grows and then it cannot be ignored when the energy density of baryon becomes comparable to that of dark matter.

Note that the Jordan frame energy density of baryon, $\rho_b$, always decays as $a_J^{-3}$ where $a_J=Aa$ is the scale factor of the Jordan frame metric. The quantity $A^4 \rho_b$ is the energy density in the Einstein frame.

In the Einstein frame, the interpretation of the peculiar behavior of $A^4\rho_G$ and $A^4\rho_b$ is that the energy density of massive gravitons is converted to that of baryon through the motion of the chameleon field $\phi$. Although we have considered the non-relativistic massive gravitons, the energy density of that in the Einstein frame behaves as radiation which implies that the field $\phi$ removes the energy of massive gravitons (indeed, the graviton mass decreases due to the motion of $\phi$). The removed energy is transferred into baryon via the non-minimal coupling.

During the scaling solution, the massive graviton never dominates over radiation because both energy densities of the massive graviton and radiation obey the same decaying law $A^4\rho_r, A^4\rho_G \propto a^4$. Hence, the field $\phi$ can reach the bottom of the effective potential before the MD era. After reaching the bottom of the effective potential, the standard decaying laws for matters $A^4\rho_r \propto a^{-4}$ and $A^4\rho_G, A^4 \rho_b \propto a^{-3}$ are recovered, then the usual dynamics of the universe is obtained with the observed dark matter-baryon ratio. 

We note that the following-up of the baryon energy density can be realized even if the scaling solution does not exist $(\lambda^2<2)$. The dynamics of this case is numerically studied in Sec.~\ref{sec_numerical}.

\subsubsection{$\rho_{b,i}\gg \rho_{G,i}$ before the equal time}
\label{b>>DM}
In this case,  the equations for the scale factor and $\phi$ form a closed system given by
\begin{align}
\ddot{\phi}+3H\dot{\phi}+\frac{\beta}{M_p} A^4 \rho_b=0\,,
\\
3M_p^2 H^2=A^4 \rho_r+ A^4 \rho_b+\frac{1}{2}\dot{\phi}^2
\,. 
\end{align}
The scaling solution is then found as
\begin{align}
\phi&=-\frac{M_p}{2\beta} \ln t +{\rm constant} \,,
\nn
a &\propto t^{1/2}
\,,
\end{align}
in which the density parameters are
\begin{align}
\Omega_b=\frac{1}{3 \beta^2}
\,, \quad
\Omega_{\phi}=\frac{1}{6\beta^2}
\,, \quad
\Omega_r=1-\frac{1}{2\beta^2}
\,,
\end{align}
where $\beta$ has to satisfy $\beta^2>1/2$.

During this scaling solution, the universe does not expand in the Jordan frame. Although the Einstein frame scale factor expands as the RD universe, $a\propto t^{1/2}$, the Jordan frame scale factor is given by
\begin{align}
a_J=aA={\rm constant}
\,.
\end{align}

The solution for $u$ is found by substituting the scaling solution into \eqref{eq_u}. We obtain
\begin{align}
u\propto a^{-\frac{3}{2}+\frac{\lambda}{4\beta}}
\,,\quad
m_T^2 \propto a^{\lambda/\beta}
\,,
\end{align}
and then the energy density of massive graviton varies as
\begin{align}
A^4 \rho_G \propto  a^{-3+\lambda/2\beta}\,.
\end{align}
The adiabatic condition \eqref{adiabatic_condition} is guaranteed when $\epsilon \ll 1$.
When $2\beta \simeq 5 \lambda$, the graviton mass roughly increases as $m_T^2 \propto a^{2/5}$ and the energy density of massive gravitons in the Einstein frame decreases as $A^4\rho_G \propto a^{-14/5}$. Therefore, even if the energy density of massive gravitons is significantly lower than that of baryon, the correct dark matter-baryon ratio is realized in time since the energy density of massive gravitons decreases slower than that of baryon.

\subsection{Matter dominant era}
After $\phi$ reaches the potential minimum $\phi_{\rm min}$, the chameleon field $\phi$ does not move during the MD universe. As shown in \cite{Aoki:2017cnz}, when $\phi$ is constant, the massive graviton behaves as CDM and then the standard MD universe is obtained.

\subsection{Accelerating expanding era}
\label{sec_acc}
After the MD era, the universe must show the accelerating expansion due to dark energy. Although one can introduce a new field to obtain the acceleration, we consider a minimal scenario such that the chameleon field itself is dark energy, i.e., the accelerating expansion is realized by the potential $V_0$. When $V_0$ becomes relevant to the dynamics of $\phi$, the chameleon field again rolls down which leads to a decreasing of $m_T$. As a result, the energy density of massive gravitons rapidly decreases and then we can ignore the contributions from massive gravitons. The basic equation during the accelerating expansion is thus given by
\begin{align}
3M_p^2 H^2=A^4 \rho_b+\frac{1}{2}\dot{\phi}^2 +V_0 
\,, \label{Fri_DE} \\
\ddot{\phi}+3H\dot{\phi}-\frac{\lambda}{M_p}V_0+\frac{\beta}{M_p}A^4 \rho_b=0
\,, \label{eq_phi_DE}
\end{align}
which yield a scaling solution
\begin{align}
\phi&=\frac{2M_p}{\lambda} \ln t +{\rm constant}
\,, \nn
a&\propto t^{\frac{2}{3}(1+\beta/\lambda)} \,,
\end{align}
in which
\begin{align}
\Omega_b=\frac{\lambda^2+\beta \lambda -3}{(\beta+\lambda)^2}
\,, \quad
\Omega_{\phi}=\frac{\beta^2+\beta \lambda +3}{(\beta+\lambda)^2}
\,,
\end{align}
and
\begin{align}
w_{\rm tot}=-\frac{2\beta}{4\beta+\lambda}
\,.
\end{align}
The scaling solution exists when
\begin{align}
\lambda(\beta+\lambda)>3
\,. \label{inequality_DE}
\end{align}
For $2\beta=5\lambda$, we find $w_{\rm tot}=-5/11$ and the inequality \eqref{inequality_DE} is reduced into $\lambda^2 >6/7$.

The amplitude of the massive graviton is given by
\begin{align}
u \propto t^{-\frac{1}{2}(1+2\beta/\lambda)}
\,,
\end{align}
and then the density parameter of massive graviton decreases as
\begin{align}
\Omega_G\propto t^{-1-2\beta/\lambda}
\,.
\end{align}
The energy density of massive graviton gives just a negligible contribution during this scaling solution which guarantees the equations \eqref{Fri_DE} and \eqref{eq_phi_DE}.

On the other hand, when $\lambda^2<6/7$, the non-minimal coupling is small so that the field $\phi$ can be approximated as a standard quintessence field. As a result, the acceleration is obtained by the slow-roll of $\phi$ and then the dark energy dominant universe is realized.


\section{Cosmic evolutions}
\label{sec_numerical}
In this section, we numerically solve the equations \eqref{Fri}-\eqref{eq_u}. We discuss two cases, the over-produced case ($\rho_{G,i} \gg \rho_{b,i}$) and the less-produced case ($\rho_{G,i}\ll \rho_{b,i}$), in order.

\subsection{Over-produced case}
\begin{figure}[tbp]
\centering
\includegraphics[width=7cm,angle=0,clip]{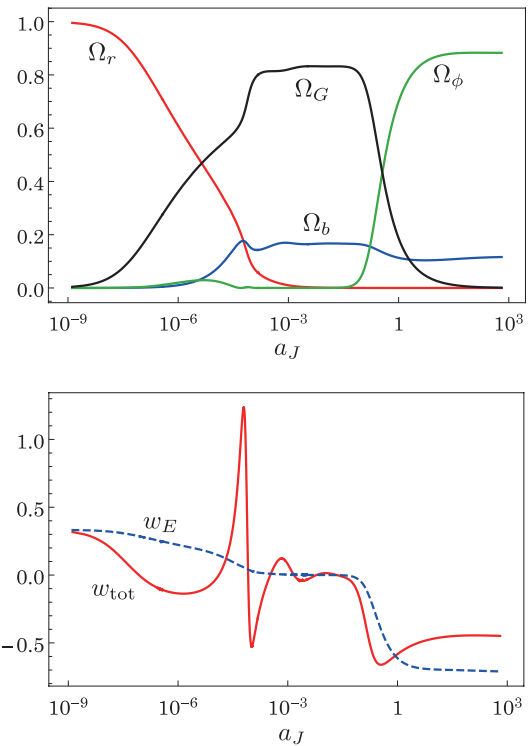}
\caption{The evolution of the density parameters and the total equation of state parameters in terms of the Jordan frame scale factor $a_J=Aa$ which is normalized to be $\Omega_{\phi}|_{a_J=1}=0.7$. We set $\beta=3$ and $\lambda=2\beta/5=6/5$ in Model A \eqref{model_A}. We assume the initial ratio between dark matter and baryon as $\rho_{G,i}/\rho_{b,i}=\Omega_{G,i}/\Omega_{b,i}=200$ with $\phi_i=0$.
}
\label{fig_modelA}
\end{figure}

First, we consider the over-produced case $\rho_{G,i} \gg \rho_{b,i}$. We assume \eqref{model_A} which we call Model A. A cosmological dynamics is shown in Fig.~\ref{fig_modelA}. We set $\rho_{G,i}/\rho_{b,i}=\Omega_{G,i}/\Omega_{b,i}=200$ at the initial of the numerical calculation. Although dark matter is initially over-produced, the energy density of baryon follows up that of dark matter and then we obtain $\rho_G/\rho_b\simeq 5$ when $a_J=Aa\sim 10^{-4}$ where we normalize the Jordan frame scale factor $a_J$ so that $\Omega_{\phi}|_{a_J=1}=0.7$. We note the following-up of $\rho_b$ is obtained even if $\lambda^2>2$ is not satisfied (In Fig.~\ref{fig_modelA}, we set $\lambda^2=(6/5)^2<2$).

\begin{figure}[tbp]
\centering
\includegraphics[width=7cm,angle=0,clip]{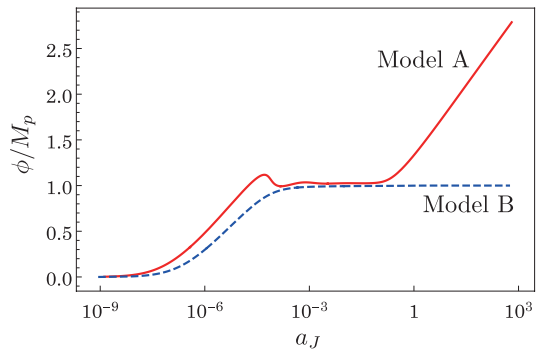}
\caption{The evolution of the chameleon field $\phi$ in Model A and Model B with $\rho_{G,i}/\rho_{b,i}=200$.
}
\label{fig_phi}
\end{figure}

The dynamics of the universe is precisely tested by the CMB observations after the decoupling time $a_J \simeq 10^{-3}$. The evolutions of the total equation of state parameters are shown in Fig.~\ref{fig_modelA}. The dynamics of the observable universe is represented by the Jordan frame quantity $w_{\rm tot}$ because the visible matters couple with the Jordan frame metric. On the other hand, since the dark matter (i.e., massive gravitons) is originated from the the gravity sector, dark matter feels the dynamics of the Einstein frame whose equation of state parameter is denoted by $w_E$. Although the large deviation of dynamics from the standard cosmological one appears before the decoupling time $a_J \lesssim 10^{-3}$, the standard dust dominant universe is recovered around the decoupling time.

When we increase the values of $\beta$ and $\lambda$, the deviation from the standard evolution is amplified which is caused by the oscillation of $\phi$ around $\phi_{\rm min}$ as shown in Fig,~\ref{fig_phi}. Since the Jordan frame scale factor is given by $a_J=Aa=a e^{\beta\phi/M_p}$, the oscillation of $\phi$ yields the oscillation of $a_J$ which is amplified by increasing of $\beta$.

Fig.~\ref{fig_modelA} does not show the dark energy ``dominant'' universe even in the accelerating phase. Instead, the acceleration is realized by the scaling solution as explained in Sec.~\ref{sec_analytic}. If this scaling solution can pass the observational constraints, it might give an answer for the other coincidence problem of dark energy: why the present dark energy density is almost same as that of matter? However, the cosmological dynamics after the decoupling time is strongly constrained by the observations. Thus, the dark energy model with the scaling solution should have a severe constraint (see \cite{Amendola:1999er,Amendola:2003eq} for examples). Furthermore, the large coupling $\alpha_A \gtrsim M_p^{-1}$ leads to that the Compton wavelength of the chameleon field has to be less than Mpc to screen the fifth force in the Solar System~\cite{Wang:2012kj}; however, the coupling functions \eqref{model_A} require the Gpc scale Compton wavelength to give the current accelerating expansion.

\begin{figure}[tbp]
\centering
\includegraphics[width=7cm,angle=0,clip]{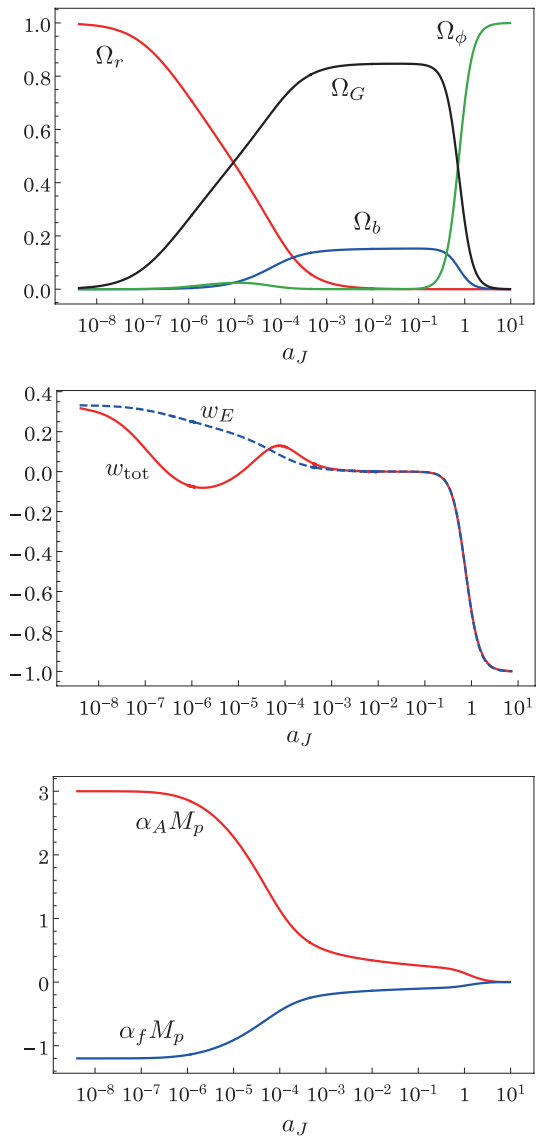}
\caption{The same figures as Fig.~\ref{fig_modelA} and the evolution of $\alpha_A$ and $\alpha_f$ in Model B. We set $\beta=3, \lambda=6/5$ and $M=M_p$. We assume the same initial condition as Fig.~\ref{fig_modelA}.
}
\label{fig_modelB}
\end{figure}

We then provide a model in which the couplings $\alpha_A$ and $\alpha_f$ are initially large but they become small in time. This behavior is realized by the model
\begin{align}
K^2=(1-\phi^2/M^2)^{-1} , \, A=e^{\beta\phi/M_p} ,\,
f=e^{-\lambda \phi/M_p}, \label{model_B}
\end{align}
which we call Model B. The only difference from Model A is that $K$ is a function of $\phi$. If the amplitude of the field $\phi$ is small at initial $(\phi \ll M)$, Model B gives the same behavior as Model A. After $\phi$ starts to roll and then $|\phi| \rightarrow M$, the kinetic function $K$ increases which causes the decreasing of the non-minimal couplings $\alpha_A,\alpha_f \rightarrow 0$ (see Figs.~\ref{fig_phi} and \ref{fig_modelB}). 
Note that the field value is restricted in the range $-M<\phi<M$ in Model B which gives a constraint $|\phi_{\rm min}|<M$ to obtain $\rho_G/\rho_b\simeq 5$.

A numerical solution of Model B is shown in Fig.~\ref{fig_modelB}. The evolutions of $\alpha_f$ and $\alpha_A$ are shown in the bottom of Fig.~\ref{fig_modelB}. As we expected, the couplings become weak in time. In particular, if $\phi_{\rm min} \simeq M$ in Model B, the cosmological dynamics is quite similar to that in the $\Lambda$CDM model after the decoupling time.

The evolutions in Model A and Model B are divided into four regimes: the radiation dominant era $(a_J \lesssim 10^{-8})$, the following-up era $(10^{-8}\lesssim a_J \lesssim 10^{-3})$, the dust dominant era $(10^{-3}\lesssim a_J \gtrsim 10^{-1})$, and the accelerating era $(10^{-1}\lesssim a_J)$. In Model A, the deviations from the $\Lambda$CDM model appear in the following-up era and the accelerating era. On the other hand, the deviation appears only in the following-up era in Model B which is after the Big Bang Nucleosynthesis (BBN) $a_J\simeq 10^{-8}$ and before the CMB last scattering surface $a_J\simeq 10^{-3}$. Since the relation between the temperature and the Hubble expansion rate during the BBN era is essentially the same as in the standard cosmology, Model B is compatible with the standard BBN.  In Model B, evolution of perturbations at the CMB scales is also expected to be the same as in the standard cosmology. On the other hand, the non-standard evolution before the CMB last scattering surface may change the evolution of perturbations at smaller scales. These deviations may give observational constraints on our models or may help addressing some of the tensions between the standard $\Lambda$CDM and observational data at small scales.

\subsection{Less-produced case}
\begin{figure}[tbp]
\centering
\includegraphics[width=7cm,angle=0,clip]{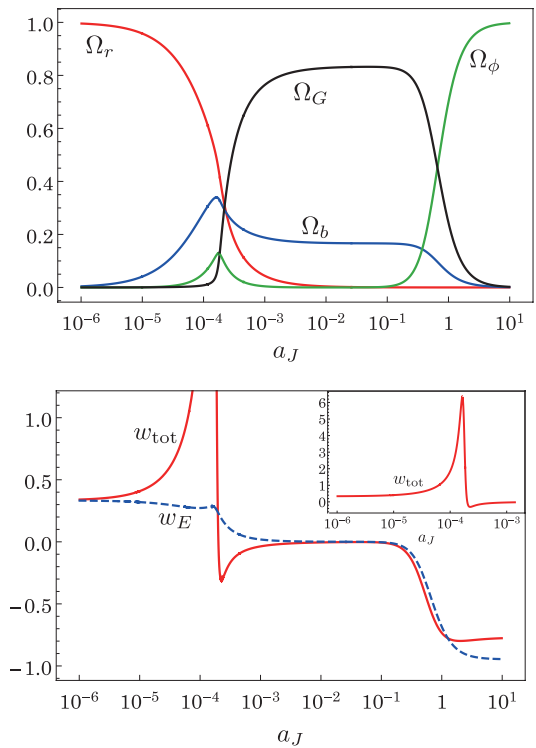}
\caption{The evolution of the density parameters and the total equation of state parameter. We set $\beta=1$ and $\lambda=2/5$ in Model A. We assume the initial ratio between dark matter and baryon as $\rho_{G,i}/\rho_{b,i}=0.01$ with $\phi_i=0$.
}
\label{fig_modelA_less}
\end{figure}

Next, we discuss the case of the less-produced dark matter $\rho_{G,i}\ll \rho_{b,i}$. A numerical solution is shown in Fig.~\ref{fig_modelA_less}. Although the energy density of dark matter is initially smaller than that of baryon, the correct abundance $(\rho_G/\rho_b\simeq 5)$ is obtained. Since the non-minimal coupling is not so large ($\beta=1$ in the case of Fig.~\ref{fig_modelA_less}), the universe evolves into the dark energy dominant universe in the future.

The following-up era is realized when $\Omega_b \simeq 1/3\beta^2$ in the scaling solution in which the dynamics of the universe is deviated from the standard one. Hence, the small value of $\beta$ yields that the following-up era is close to the decoupling time and then the deviation may give a large effect on the CMB physics. For instance, Fig.~\ref{fig_modelA_less} indicates that the deviation still exists at $a_J \simeq 10^{-3}$.

\begin{figure}[tbp]
\centering
\includegraphics[width=7cm,angle=0,clip]{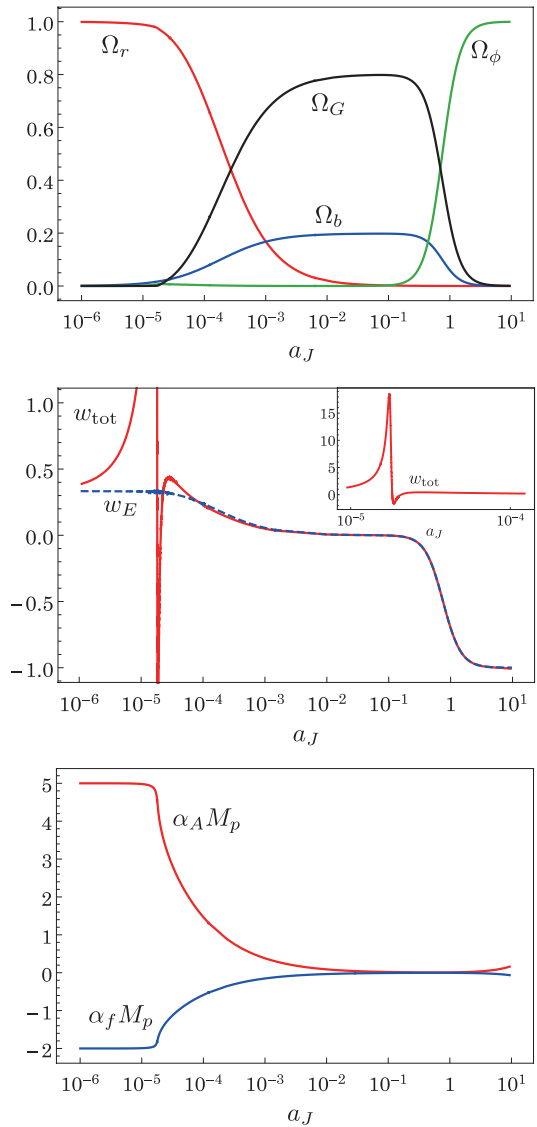}
\caption{The same figures as Fig.~\ref{fig_modelB} in the less-produced case. We set $\beta=5, \lambda=2$ and $M=M_p$, and assume $\rho_{G,i}/\rho_{b,i}=0.01$ with $\phi_i=0$.
}
\label{fig_modelB_less}
\end{figure}

\begin{figure}[tbp]
\centering
\includegraphics[width=7cm,angle=0,clip]{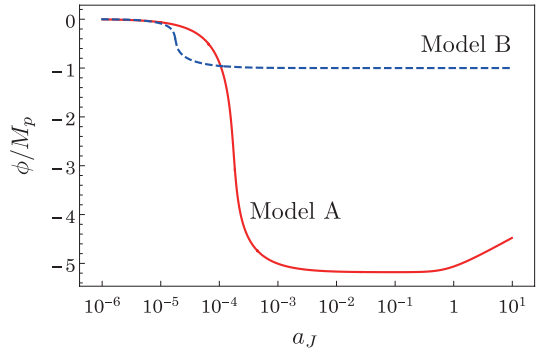}
\caption{The evolution of the chameleon field $\phi$ with $\rho_{G,i}/\rho_{b,i}=0.01$ and $\phi_i=0$. We set $\beta=1, \lambda=2/5$ for Model A and $\beta=5, \lambda=2, M=M_p$ for Model B.
}
\label{fig_phi_less}
\end{figure}

In Model B, the coupling strength is time-dependent. Hence, we can obtain a scenario in which the non-minimal couplings are initially large whereas the couplings turn to be weak after the ratio $\rho_G/\rho_b$ is dynamically adjusted to the observed value as shown in Fig.~\ref{fig_modelB_less}. In this case, the standard cosmological dynamics is recovered around $a_J\simeq 10^{-4}$. Although the dynamics is deviated from the standard one in $10^{-6} \lesssim a_J \lesssim 10^{-4}$, it should have little effect on CMB.

When the bare potential $V_0$ becomes relevant to $\phi$, the field value increases as shown in Fig.~\ref{fig_phi_less}. In Model B, this increasing leads to the increasing of the non-minimal couplings $\alpha_A$ and $\alpha_f$ in the future $a_J \gg 1$. Needless to say, if the form of the bare potential $V_0$ is modified in order that $\dot{\phi}<0$ in the dark energy dominant era (for example, $\lambda<0$), the coupling will be weak in the future as with Fig.~\ref{fig_modelB}.

As a result, we can obtain a viable cosmological dynamics even in the less-produced case. Although the initial abundance of dark matter is much smaller than the observed value, the chameleon field provides the energy transfer from baryon to dark matter via the non-minimal couplings. The correct abundance of dark matter can be realized without the fine-tuning of the initial condition.


\section{Concluding remarks}
\label{summary}
In the present paper, we provide a cosmological scenario by which the observed dark matter-baryon ratio can be naturally explained.  We have added two new ingredients to the standard model: the massive graviton and the chameleon field corresponding to dark matter and dark energy, respectively. The matter fields are minimally coupled to the Jordan frame metric which leads to the non-minimal coupling of the matter fields to the chameleon field. On the other hand, the chameleon field may have a different coupling to dark matter since dark matter, the massive graviton, is originated from the gravity sector. We have assumed that the chameleon field has a non-minimal coupling to the massive graviton via the mass terms of the graviton. Two different non-minimal couplings of the chameleon field realize that the ratio $\rho_G/\rho_b$ is dynamically relaxed to $\sim 5$ without the tuning of the initial condition. 

We have studied two simple models: Eq.~\eqref{model_A} (Model A) and Eq.~\eqref{model_B} (Model B).\footnote{Although we chose exponential forms of $A$ and $f$, other types of the non-minimal coupling can be discussed. Even in this case, the dark matter-baryon ratio could be explained when $\alpha_A/\alpha_f \simeq -5/2$.} In these models, we have an additional era in the universe before the radiation-matter equality which we have called the following-up era. Even if dark matter is initially over-produced or less-produced, the chameleon field transfers the energy of the larger one of dark matter and baryon into that of the smaller one via the non-minimal couplings. After the ratio $\rho_G/\rho_b$ is tuned, the standard cosmological dynamics can be recovered.

To realize the following-up era, the non-minimal couplings should be at least comparable to the usual gravitational interaction, $|\alpha_A|, |\alpha_f| \sim M_p^{-1}$. This ``large'' non-minimal couplings may lead to a deviation from the $\Lambda$CDM cosmology. However, in Model B, the coupling strengths are time-dependent and then the present non-minimal couplings can be small as shown in Fig.~\ref{fig_modelB} and Fig.~\ref{fig_modelB_less}.

The chameleon bigravity theory \eqref{action} was originally proposed in \cite{DeFelice:2017oym} in order to avoid the Higuchi instability of cosmological solutions~\cite{Higuchi:1986py,Higuchi:1989gz,Grisa:2009yy,Fasiello:2012rw,Fasiello:2013woa,Comelli:2012db,Comelli:2014bqa,DeFelice:2014nja,Aoki:2015xqa} and to circumvent the low cutoff scale of the standard bigravity theory. When the graviton mass is smaller than the Hubble expansion rate, the scalar mode of the massive graviton exhibits the ghost instability or the gradient instability, in general. However, the chameleon bigravity can yield a cosmological solution in which $m_T/H$ is constant during the radiation dominant universe and then the homogeneous and isotropic solution does not suffer from the instability even in the early universe~\cite{DeFelice:2017oym,DeFelice:2017gzc}. The cutoff scale of the gravity sector also becomes high in the early universe and thus this mechanism make it possible for us to apply the bigravity theory to the early universe. Since we have considered only the cases with $m_T \gg H$, the Higuchi instability is not problematic for the present discussion and thus we have not discussed the stability issue of the homogeneous spacetimes \eqref{Bianchi_g} and \eqref{Bianchi_f}. Nevertheless, it would be interesting to study whether or not the chameleon bigravity also gives a stable homogeneous but anisotropic solution even in the early epoch of the universe.

As discussed in \cite{Aoki:2016zgp,Aoki:2017cnz}, the massive graviton with a constant mass is a viable dark matter candidate in the wide range of the mass $10^{-23}{\rm eV}\lesssim m \lesssim 10^7 (M_g/M_f)^{2/3} {\rm eV}$. The lower bound is imposed so that dark matter halos form in dwarf galaxy scales, while the upper bound is given by the requirement that the lifetime of the massive graviton be longer than the age of the universe. In the chameleon bigravity theory studied in the present paper, by the same argument as before, we obtain the lower bound on the present value of $m_T$, i.e., $m_T|_{a_J = 1} \gtrsim 10^{-23}{\rm eV}$. The lifetime of the massive graviton also constrains the present value of the mass as follows. Since the abundance of dark matter is automatically tuned to the observed value  by the scaling solution, we need to discuss the lifetime of the massive graviton only after the following-up era in our scenario~\footnote{Decay of massive graviton during and after the nucleosynthesis but before the end of the following-up era may have some impacts on the primordial abundances of the light elements. We leave studies of such effects to future publications.}. The graviton mass remains almost constant during this epoch. As a result, the requirement of a long enough lifetime gives the upper bound on the present value of the mass as $m_T|_{a_J=1} \lesssim 10^7 (M_g/M_f)^{2/3} {\rm eV}$.

Since we have not discussed concrete observational constraints on our models, the observational viability of the models is an open question. Theoretically, our model can explain the dark matter-baryon ratio for any initial condition except for $\rho_{G,i}=0$ or $\rho_{b,i}=0$ because the following-up era can be the scaling solution and then it never ends unless the smaller one of $\rho_g$ and $\rho_b$ catches up the larger one. However, for example, $\rho_{G,i}/\rho_{b,i} \simeq 200$ should be an observational upper bound on the initial ratio in the over-produced case with $\beta=3,\lambda=6/5$. If $\rho_{G,i}/\rho_{b,i} \gtrsim 200$, the dynamics of the universe at BBN is changed from the radiation dominant universe and then it may give an observational constraint. Furthermore, we should take into account local gravity constraints of the fifth force since the following-up of dark matter or baryon requires a large non-minimal coupling of the chameleon field. Although the current non-minimal coupling can be small in Model B, there still exists a small fifth force and then the local gravity experiments may give a constraint on our model as well.\footnote{The massive graviton also yields a fifth force. We should also discuss the Vainshtein screening mechanism of the massive graviton. However, if the graviton mass is high enough (e.g., $m_T|_{a_J=1} \gg 10^{-4}$ eV), the fifth force propagated by the massive graviton does not exist in the observed scales due to the Yukawa suppression.} We leave the details of the observational constraints of our models for a future work.

\section*{Acknowledgments}
The work of K.A. was supported in part by Grants-in-Aid from the Scientific Research Fund of the Japan Society for the Promotion of Science  (No. 15J05540). 
The work of S.M. was supported in part by JSPS Grant-in-Aid for Scientific Research No. 17H02890, No. 17H06359,  and by World Premier International Research Center Initiative (WPI), MEXT, Japan.


\appendix

\section{Bianchi I universe}
We consider the axisymmetric Bianchi I universe \eqref{Bianchi_g} and \eqref{Bianchi_f}. We consider the matter field minimally coupled with $\tilde{g}_{\mu\nu}$ whose energy-momentum tensor is given by the form \eqref{Tmunu} and the conservation law is given by \eqref{conservation}.
We find following equations: the Friedmann equations
\begin{widetext}
\begin{align}
3H^2&=\frac{1}{M_g^2} \left[ \frac{1}{2}K^2\dot{\phi}^2+\rho A^4\right] +3\Sigma_g^2+m^2 f(\phi) [c_0+c_1(e^{-2 \sigma }+2e^{\sigma}) \xi +c_2(2e^{-\sigma}+e^{2\sigma}) \xi^2 + c_3 \xi^3 ]
\,, \label{g_Fri}\\
3\hat{H}^2&=3 \Sigma_f^2+\frac{m^2}{\kappa}f(\phi)[c_4 +c_3(2e^{-\sigma}+e^{2\sigma} ) \xi^{-1} + c_2 (e^{-2\sigma}+2 e^{\sigma}) \xi^{-2}+c_1 \xi^{-3}]
\,, \label{f_Fri}
\end{align}
the equations for the shears
\begin{align}
\frac{3}{a^3} \frac{d}{dt}\left[ a^3  \Sigma_g\right]-m^2 f(\phi) (e^{\sigma}-e^{-2\sigma})(c_1\xi +c_2(c+e^{\sigma})\xi+c_3 c e^{\sigma}\xi^3 )&=0\,,
\label{eq_sigmag}
\\
\frac{3}{a^3} \frac{d}{dt}\left[ (\xi a)^3 \Sigma_f \right] +\frac{m^2}{\kappa} f(\phi) (e^{\sigma}-e^{-2\sigma})(c_1\xi +c_2(c+e^{\sigma})\xi+c_3 c e^{\sigma}\xi^3 )&=0
\label{eq_sigmaf}\,,
\end{align}
the equation for the chameleon field
\begin{align}
K\left( \ddot{\phi}+3H\dot{\phi}\right)+\dot{K}\dot{\phi}+\alpha_f m^2M_g^2f(\phi)
\Bigl[& c_0+c_1(2e^{\sigma}+e^{-2\sigma}+c)\xi
+c_2\{ 2e^{-\sigma}+e^{2\sigma}+c(1+2e^{\sigma})\}\xi^2
\nn
&+c_3\{ 1+c(2e^{-\sigma}+e^{2\sigma}) \} \xi^3
+c_4 c\xi^4 \Bigl]
+\alpha_A A^4 (\rho-3P)=0\,,
\end{align}
and the constraint
\begin{align}
&H \left[ 3c_1+2c_2\xi(2e^{\sigma}+e^{-2\sigma})+c_3\xi^2 (e^{2\sigma}+2e^{-\sigma}) \right]
\nn
-&\hat{H} \xi \left[ 3c_3 \xi^2 +2c_2\xi (e^{2\sigma}+2e^{-\sigma})+c_1(2e^{\sigma}+e^{-2\sigma}) \right]
\nn
+&2\xi (e^{-\sigma}-e^{2\sigma})\left[ \Sigma_f(c_1e^{-\sigma}+c_2 \xi)+\Sigma_g(c_2 e^{-\sigma}+c_3 \xi) \right]
\nn
+&\alpha_f K \dot{\phi}\xi^3
\left[ c_4+c_3(2e^{-\sigma}+e^{2\sigma})\xi^{-1} c_2 (e^{-2\sigma}+2e^{\sigma})\xi^{-2} +c_1 \xi^{-3} \right]  
=0
\label{BianchiI_constraint}
\,,
\end{align}
\end{widetext}
where we have defined
\begin{align}
H& :=\frac{\dot{a}}{a} \,, \quad \hat{H}:= \frac{(\xi a)^{\cdot}}{ac\xi^2 } \,, \\
\Sigma_g& := \dot{\sigma}_g\,, \quad \Sigma_f:= \frac{\dot{\sigma}_f}{c\xi}\,,
\\
\sigma& :=\sigma_g-\sigma_f\,.
\end{align}

We then expand the equations in terms of $\sigma$. Note that when the anisotropies are dominant components of the universe the amplitude of $\sigma$ is given by
\begin{align}
\sigma \sim \epsilon \,,
\end{align}
with $\epsilon =H/m_T$.
Hence, the small anisotropy can be the dominant component when the graviton mass is larger than the Hubble expansion rate $(\epsilon \ll 1)$. 
Note that the assumption $\epsilon \ll 1$ leads to
\begin{align}
c_0+3c_1+3c_2+c_3 \lesssim \epsilon^2 \,, \label{small_V}
\end{align}
from the consistency of the Friedmann equation which is obtained from the condition $V_0 \leq 3M_p^2 H^2$ (see Eq.~\eqref{Friedmann_eq}).

At the stage of the universe with $\epsilon \ll 1$, the spacetimes evolve to be $\xi \rightarrow \xi_c$ and $c\rightarrow 1$ with a constant $\xi_c$ where $\xi_c$ is a root of the algebraic equation
\begin{align}
&\frac{1}{M_g^2}(c_0+3c_1\xi_c +3c_2 \xi_c^2 +c_3 \xi_c^3)
\nn
=\, &
\frac{1}{\xi_c^2M_f^2}(c_1 \xi_c+3c_2 \xi_c^3+3c_3 \xi_c^3 +c_4\xi_c^4) \,.
\end{align}
By rescaling the coupling constant such that
\begin{align}
c_i \rightarrow \xi_c^{-i} c_i\,, \quad M_f \rightarrow \xi_c^{-1} M_f\,, 
\end{align} 
we can always set $\xi_c=1$ in which the (rescaled) coupling constants satisfy
\begin{align}
\kappa (c_0+3c_1+3c_2+c_3)=(c_1+3c_2+3c_3+c_4)\,.
\end{align}
We can also expand the equations in terms of $\delta \xi$ and $\delta c$ defined by
\begin{align}
\delta \xi=\xi-1\,, \quad \delta c =c-1 \,.
\end{align}

The constraint equation yields
\begin{align}
&3(H-\hat{H})(c_1+2c_2+c_3)
\nn
+\, &\alpha_f (c_1+3c_2+3c_3+c_4)K\dot{\phi}=0+\mathcal{O}(H\epsilon)
\,.
\end{align}
The inequality \eqref{small_V} reads
\begin{align}
\hat{H}=H+\mathcal{O}(H\epsilon)\,.
\end{align}
The Friedmann equations \eqref{g_Fri} and \eqref{f_Fri} then give
\begin{align}
&m_T^2\delta \xi + \frac{1}{3M_g^2}\left[ K^2\dot{\phi}^2+\rho A^4 \right]+\Sigma_g^2-\Sigma_f^2
\nn
+\,&m^2f(\phi) \left[ (c_1+c_2)-\kappa^{-1}(c_2+c_3) \right]\sigma^2
\nn
=&\, 0
+\mathcal{O}(H^2\epsilon)
\,. \label{delta_xi}
\end{align}
Solving this equation with respect to $\delta \xi$ and substituting it into Eq.~\eqref{g_Fri} we find
\begin{align}
3M_p^2H^2= 
&\, \rho A^4+\frac{1}{2}K^2\dot{\phi}^2+V_0+\frac{1}{2}\dot{h}^2
\nn
&+\frac{1}{2}\dot{\varphi}^2+\frac{1}{2}m_T^2\varphi^2 
+\mathcal{O}(M_p^2H^2 \epsilon)\,,
\label{Friedmann_eq}
\end{align}
where we have introduced the normalized mass eigenstate $\varphi$ and $h$ defined by the relations
\begin{align}
\sigma_g&=\frac{1}{\sqrt{6} M_p}\left (h+\kappa^{1/2}\varphi \right) \,, \label{graviton1} \\
\sigma_f&=\frac{1}{\sqrt{6} M_p}\left( h-\kappa^{-1/2}\varphi \right)  \,. \label{graviton2}
\end{align}
The variable $\delta c$ is determined by 
\begin{align}
\delta c=\frac{1}{\hat{H}}\left(\frac{\dot{\xi}}{\xi}+H \right)-1\,. \label{delta_c}
\end{align}
One can estimate the typical amplitudes of $\delta \xi$ and $\delta c$ from Eq.~\eqref{delta_xi} and Eq.~\eqref{delta_c}, which are
\begin{align}
\delta \xi\sim \epsilon^2\,,\quad \delta c \sim \epsilon^2\,.
\end{align}
Then the equations of the anisotropies and the chameleon field are reduced to Eq.~\eqref{eq_varphi}, Eq.~\eqref{eq_cham}, and
\begin{align}
\ddot{h}+3H\dot{h}&=0 \,,
\end{align}
The equation of $h$ reads $\dot{h}\propto a^{-3}$ and then its contribution to the Friedmann equation \eqref{Friedmann_eq} decreases as $a^{-6}$ which is faster than others. Hence, we can ignore $h$ in time and obtain Eq.~\eqref{Fri_no_h}.

\bibliography{ref}
\bibliographystyle{JHEP}

\end{document}